\documentclass[letter,traditabstract]{aa}

\usepackage{natbib}
\usepackage{dcolumn,lscape,longtable} 
\bibpunct{(}{)}{;}{a}{}{,}
\usepackage{graphicx}
\usepackage{txfonts}

\newcommand{\cb}{CB\,244}
\newcommand{\spitzer}{\textit{Spitzer}}
\newcommand{\herschel}{\textit{Herschel}}
\newcommand \msun{\hbox{$\hbox{M}_{\odot}$}}
\newcommand \lsun{\hbox{$\hbox{L}_{\odot}$}}
\newcommand{\micron}{$\mu$m}

\begin{document}

\title{Dust--temperature of an isolated star-forming cloud:\\
  \textit{Herschel} observations of the Bok globule
  CB244\thanks{Herschel is an ESA space observatory with science
    instruments provided by European-led Principal Investigator
    consortia and with important participation from NASA.}}

   \author{A.~Stutz\inst{1,2}\and
     R.~Launhardt\inst{1}\and
     H.~Linz\inst{1}\and
     O.~Krause\inst{1}\and
     T.~Henning\inst{1}\and
     J.~Kainulainen\inst{1}\and
     M.~Nielbock\inst{1}
     J.~Steinacker\inst{3,1}\and
     P.~Andr\'e\inst{4}}

   \institute{
     Max-Planck-Institut f\"ur Astronomie, 
     K\"onigstuhl 17, D-69117 Heidelberg, Germany \and
     Department of Astronomy and Steward Observatory, 
     University of Arizona, 933 North Cherry Avenue, 
     Tucson, AZ 85721, USA \and
     LERMA \& UMR 8112 du CNRS, Observatoire de Paris, 
     61 Av. de l'Observatoire, 75014 Paris, France\and
     Laboratoire AIM, CEA/DSM-CNRS-Université Paris Diderot, 
     IRFU/Service d'Astrophysique, C.E. Saclay, Orme des 
     Merisiers, 91191 Gif-sur-Yvette, France}

   \date{Received; accepted}

\abstract{We present \herschel\ observations of the isolated,
  low--mass star--forming Bok globule CB244.  It contains two cold
  sources, a low--mass Class 0 protostar and a starless core, which is
  likely to be prestellar in nature, separated by 90$\arcsec$
  ($\sim\!18000$~AU).  The \herschel\ data sample the peak of the
  Planck spectrum for these sources, and are therefore ideal for
  dust--temperature and column density modeling.  With these data and
  a near-IR extinction map, the MIPS 70~\micron\ mosaic, the SCUBA
  850~\micron\ map, and the IRAM 1.3~mm map, we model the
  dust--temperature and column density of \cb\ and present the first
  measured dust--temperature map of an entire star--forming molecular
  cloud.  We find that the column--averaged dust--temperature near the
  protostar is $\sim\!17.7$~K, while for the starless core it is
  $\sim\!10.6$~K, and that the effect of external heating causes the
  cloud dust--temperature to rise to $\sim\!17$~K where the hydrogen
  column density drops below $10^{21}$~cm$^{-2}$.  The total hydrogen
  mass of CB\,244 (assuming a distance of 200~pc) is $15\pm5$~\msun.
  The mass of the protostellar core is $1.6\pm0.1$~\msun\ and the mass
  of the starless core is $5\pm2$~\msun, indicating that $\sim$45\% of
  the mass in the globule is participating in the star--formation
  process.}

   \keywords{globules -- ISM: individual (CB244) -- infrared: ISM
      -- (ISM:) dust, extinction } 

   \titlerunning{Dust--temperature Structure of \cb}

\maketitle

\section{Introduction}

Bok globules are nearby, small, relatively isolated molecular clouds
undergoing low--mass star--formation
\citep[e.g.,][]{clemens88,laun97}.  They tend to have only one or two
star--forming cores that are embedded in a larger common cloud.  These
relatively simple characteristics make these objects ideal for
studying the detailed processes taking place in low-mass star
formation.  Specifically, the temperature and density structure are
fundamental physical parameters necessary to understand core
fragmentation, collapse, and chemical evolution
\citep[e.g.,][]{wt07,stutz08,stutz09,laun10}.  The {\it Herschel}
\citep{pil10} data cover the wavelength range which samples the peak
of the Planck spectrum for cold sources (6 - 20~K).  This wavelength
regime is critical for accurate modeling of the temperature and
density structure in the cold environments where stars are born
\citep[e.g.,][]{shetty09b}.

The {\it Herschel} Guaranteed Time Key Program ``Earliest Phases of
Star--formation'' (EPoS; P.I.\ O. Krause) sample consists of the
Photodetector Array Camera and Spectrometer \citep[PACS;][]{pog10} and
the Spectral and Photometric Imaging Receiver
\citep[SPIRE;][]{griff10} imaging--mode observations of sites of both
high-- and low--mass star formation.  The low--mass Science
Demonstration Phase portion of the sample is the focus of this
contribution.  The source \cb\ \citep[L1262;][]{lynds62} is a Bok
globule at a distance of $\sim\!200$~pc \citep{hilton95}, with an
approximate extent of $\sim\!6\arcmin$, or about 0.5~pc.  The
\cb\ globule contains two submm {peaks, one associated with a Class 0
  protostar located at R.A.\ = $23^h 25^m 46.3^s$, Decl.\ = $+74\degr
  17\arcmin 39.1\arcsec$, and one associated with a starless core
  located at R.A.\ = $23^h 25^m 27.1^s$, Decl.\ = $+74\degr 18\arcmin
  25.3\arcsec$.  The protostar drives a molecular outflow
  \citep[e.g.,][]{clemens91}, while the detection of the \cb\ starless
  core was first published as an additional source inside the globule
  by \citet{laun96} and \citet{shirley00} and produces both an
  8~\micron\ \citep{tobin10} and a 24~\micron\ shadow.  The YSO and
  starless core are separated by $\sim\!90\arcsec$, and are therefore
  well resolved throughout the PACS and SPIRE bands.  We use
  \herschel\ imaging to construct spatially resolved spectral energy
  distributions (SEDs) of the entire cloud that cover both sides of
  the peak of the SED. These data allow us to reconstruct the
  dust--temperature map and column density distribution of the Bok
  globule and hence the density profiles and mass distribution with
  unprecedented accuracy. They also reveal the role of external
  heating and shielding by the envelopes in the energy balance of such
  cores in isolated globules.

\begin{figure*}
\begin{center}
  \scalebox{0.7}{{\includegraphics{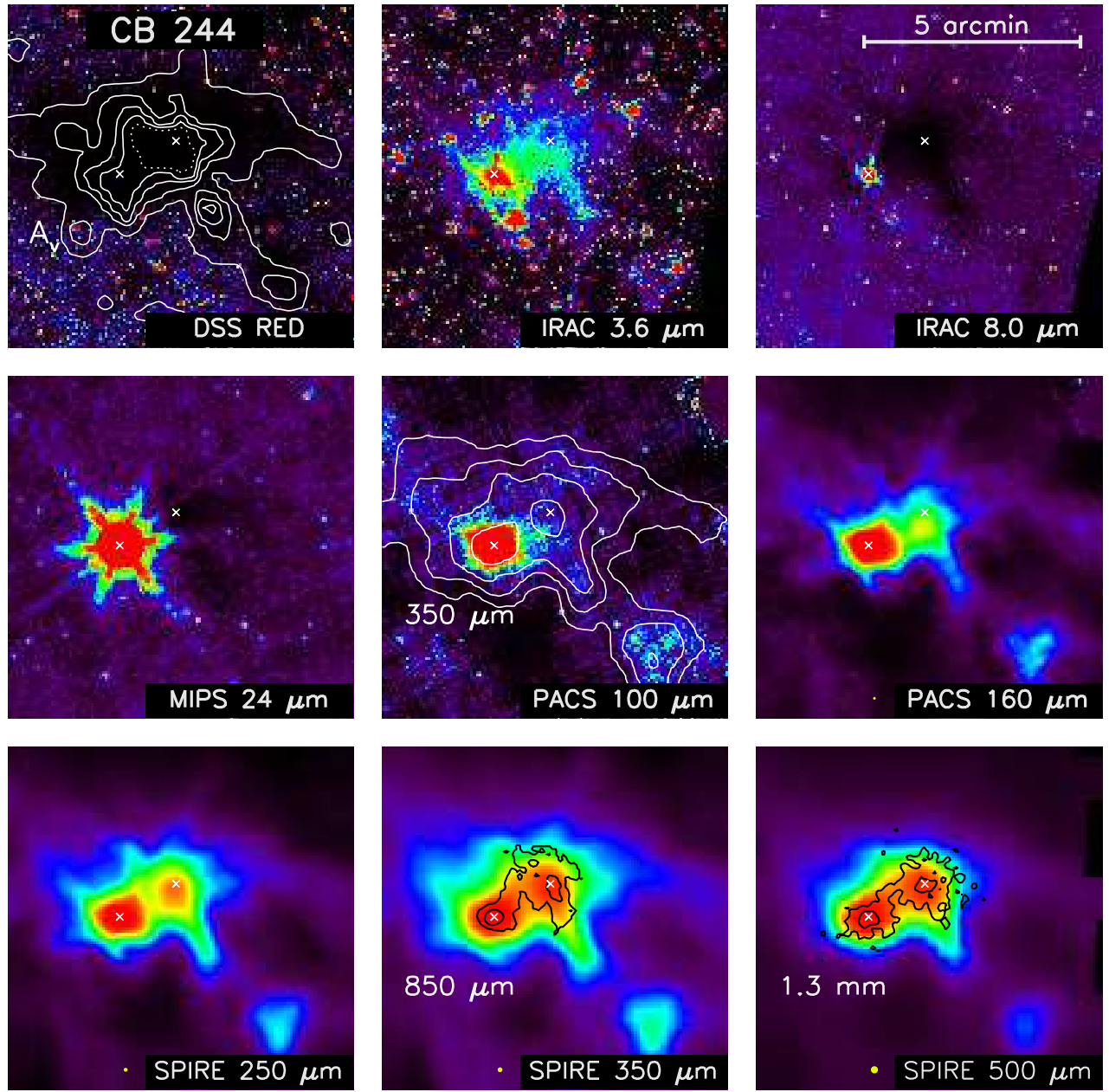}}}
  \caption{$8\arcmin \times 8\arcmin$ images of \cb, shown at the
    wavelengths indicated in the bottom--right of each panel.  The
    images are shown on a log scale at stretches intended to highlight
    the structure of absorption or emission at each wavelength.  The
    locations of the protostar and starless core, derived from the
    8~\micron\ image, are indicated with $\times$--symbols (note the
    offset between the $\times$'s and the starless core emission at
    100 -- 250~\micron\ wavelengths).  The panel contours `are labeled
    with their corresponding wavelengths.  The contour levels are:
    $A_V$ extinction map levels = \{5,10,15,20,25\}~mags (inner dotted
    contours indicates the region inside which we place lower--limits
    on the extinction); SPIRE 350~\micron\ =
    \{0.5,1.0,1.5,3.5\}~mJy/$\sq\arcsec$; SCUBA 850~\micron\ =
    \{0.3,0.75\}~mJy/$\sq\arcsec$; IRAM 1.3~mm =
    \{0.1,0.3\}~mJy/$\sq\arcsec$.  Approximate beam sizes are
    indicated as yellow circles.
    \label{fig:img}}
\end{center}
\end{figure*}

\section{Observations and data processing}

\subsection{{\textit{Herschel} observations}}

The source CB\,244 was observed with the PACS instrument on board the {\it
  Herschel} Space Observatory on 2009, December 30, during the Science
Demonstration Program.  The globule \cb\ was observed at 100~\micron\ and
160~\micron.  We obtained two orthogonal scan maps with scan leg
lengths of 9$\arcmin$ using a scan speed of 20$\arcsec$/s.  The scan
leg position angles guarantee an almost homogeneous coverage of \cb.
We produced highpass--filtered intensity maps from these data
using the HIPE software package \citep{ott10}, version 3.0, build 455.
Besides the standard steps leading to level--1 calibrated data, a
second--level deglitching as well as a correction for offsets in the
detector sub--matrices were performed. Finally, the data were 
highpass--filtered, using a median window with a width of 271 data
samples to remove the effects of bolometer temperature
drifts during the course of the data acquisition.  Furthermore, we masked
emission structures, in this case both the YSO and starless core
regions, before computing the running median.  Masking bright sources
minimizes over--subtraction of source emission in the highpass
filtering step.  Finally, the data were projected onto a coordinate
grid using the photProject routine inside HIPE. As a last step, the
flux correction factors provided by the PACS ICC team were
applied.  We note that these data have relatively flat background
values, indicating that the highpass reduction used in conjunction
with the masking is a relatively robust scheme for avoiding artifacts
and recovering extended emission in the PACS maps.  In addition to
these processing steps, we checked the pointing in the PACS
100~\micron\ map against the MIPS 24~\micron\ mosaic of the same
region.  Using point--sources detected in both images, we found a
pointing offset of $\sim\!1.25\arcsec$ in the 100~\micron\ map
relative to the 24~\micron\ mosaic.  Because the 160~\micron\ PACS map
contains no point--sources and was acquired at the same time as the
100~\micron\ data, we blindly applied the same pointing correction.  The
final PACS 100 and 160~\micron\ images are shown in
Fig.~\ref{fig:img}.

Maps at 250, 350, and 500 $\mu$m were obtained with SPIRE on October
20, 2009.  Two 9$\arcmin$ scan legs were used to cover the source. Two
repetitions resulted in 146~s of scanning time with the nominal speed
of 30$''$/s. The data were processed within HIPE with the standard
photometer script up to level 1.  During baseline removal, we masked
out the high--emission area in the center of the field.  For these
data no cross-scan was obtained; therefore, the resulting maps still
showed residual stripes along the scan direction.  We used the
\citet{bendo10} de--stripping scheme to mitigate this effect.  In
addition, we checked the pointing in the SPIRE data by
cross--correlating the images with the PACS 160~\micron\ image; we
found that the pointing offsets for all three wavelengths are on order
of 2$\arcsec$ or smaller, and therefore we did not correct for this.
The SPIRE 250, 350, and 500~\micron\ images are shown in
Fig.~\ref{fig:img}.

\subsection{Other data}

\noindent {\it Near-IR data:} We observed \cb\ in the near-IR $K_S$ and $H$
bands using Omega2000 at the Calar Alto Observatory.  The $15\arcmin
\times15\arcmin$ field--of--view provided complete coverage of the
globule.  A standard near--IR data reduction scheme was adopted.  The
extracted photometric data where calibrated using the Two Micron All
Sky Survey \citep[2MASS; ][]{skrut06}.  Following the NICE method
\citep{lada94}, the observed $(H-K_S)$ colors of stars were then
related to visual extinction by $(H-K_S) = \langle (H-K_S)_0\rangle +
0.063 \times A_V$, assuming the \citet{rieke85} extinction law, where
$\langle (H-K_S)_0\rangle$ is the mean intrinsic color of the stars
determined from edges of the field.  The $A_V$ values for each star
were used to create an extinction map.  The resulting extinction map
probes extinction values up to $A_V \approx 30$ mag, with a $3\sigma$
error of about 3 mag at $A_V=0$ mag.\\
\noindent {\it Spitzer data:} The \spitzer\ observations presented
here are from two programs: the MIPS observations are from program 53
(P.I.\ G.\ Rieke), while the IRAC observations are from program 58
(P.I.\ C.\ Lawrence).  The MIPS observations were reduced using the
Data Analysis Tool \citep[DAT;][]{gordon05} according to steps
outlined in \citet{stutz07}.  The IRAC frames were processed using the
IRAC Pipeline v14.0, and mosaics were created from the basic
calibrated data (BCD) frames using a custom IDL program
\citep[see ][]{guter08}.\\
\noindent {\it (Sub)mm continuum data:} The SCUBA 850~\micron\ and the
IRAM 1.3~mm maps are presented in \citet{laun10}.  The
850~\micron\ and 1.3~mm maps cover both sources, but do not extend
very far into the envelope.  The maps have native beam sizes of
14.9$\arcsec$ (850~\micron), and 10.9$\arcsec$ (1.3~mm). For further
details see \citet{laun10}.  In Fig.~\ref{fig:img} we show the
850~\micron\ and 1.3~mm emission as contours.

\begin{figure*}
\begin{center}
  \scalebox{0.58}{{\includegraphics[angle=-90]{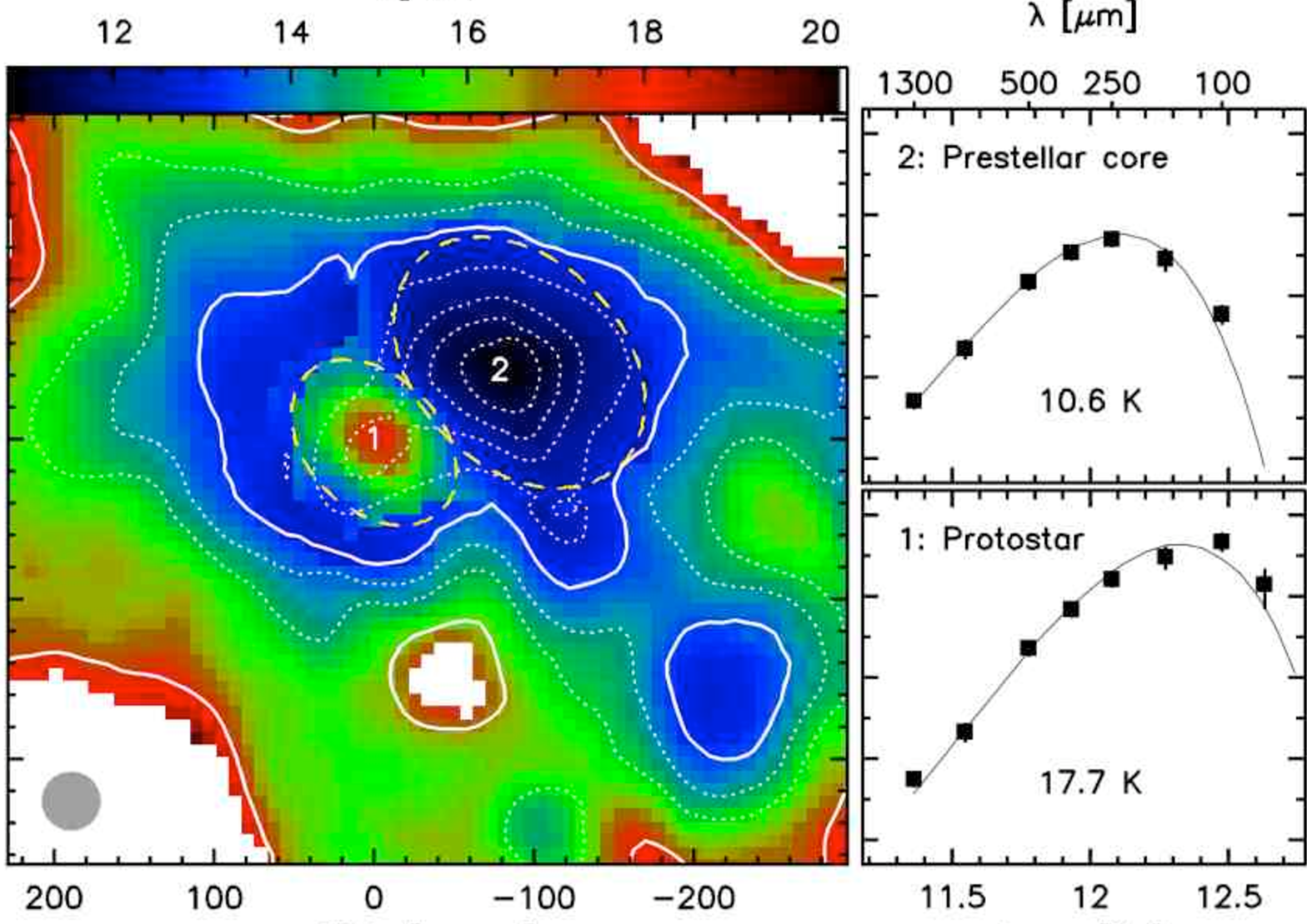}}}
  \caption{Dust--temperature (color) and hydrogen column density
    (white contours) in CB\,244. The line--of--sight--averaged mean
    dust--temperature was derived pixel by pixel from modified
    black--body SED fits to homogeneously beam--smoothed emission maps
    from \herschel\ PACS and SPIRE, \spitzer\ MIPS70, and
    ground--based extinction mapping and submillimeter dust--continuum
    maps at 0.8 and 1.3\,mm (see text for details).  The top color bar
    indicates the temperature scale.  Column density contours are at
    \{0.1 (thick) 0.3, 0.5, 1 (thick), 2, 3.5, 5, and 7\}$\times
    10^{22}$~H/cm$^2$. The peak column density in the prestellar core
    just reaches $10^{23}$ H/cm$^2$.  Two SEDs and the respective
    modified black--body fits are shown for the positions of the
    Class\,0 protostar (``1'') and the prestellar core (``2'').  The
    yellow dashed ellipses indicate the areas used to calculate source
    masses (see Sect.~4).  The 37$\arcsec$ beam size is indicated by a
    grey circle.  White areas represent regions where there is not
    enough signal--to--noise and/or coverage to get a reliable fit.
    \label{fig:temp}}
\end{center}
\end{figure*}

\section{Dust temperature and column density modeling}

We used the PACS 100 and 160~\micron, SPIRE 250, 350, and
500~\micron\ data in conjunction with the NIR extinction map, MIPS
70~\micron, SCUBA 850~\micron, and IRAM 1.3~mm dust emission
maps to model the line--of--sight--averaged temperature and column
density of \cb.  The calibrated maps were all homogeneously convolved
to the largest FWHM beam size--of the data set, the SPIRE
500~\micron\ 37$\arcsec$ beam, assuming Gaussian beam profiles, and
projected onto a common 8$\arcsec {\mathrm pix}^{-1}$\ grid. This
approach has the drawback that all spatial structures smaller than
this beam size are smoothed, but ensures that the fluxes at all
wavelengths in a given image pixel are derived from the same area in
the source.  We corrected the filtering of large spatial--scale
emission in the PACS data using the \citet{schlegel98}
100~\micron\ {\it IRAS} maps and the {\it ISO} Serendipity Survey
observations at 170~\micron: we added a zero level of emission of
0.3 and 0.8 mJy/$\sq\arcsec$ to the 100 and 160~\micron\ maps,
respectively.  

For each image pixel, an SED was extracted and fitted with a
single--temperature modified black--body of the form \mbox{$S_{\nu} =
  \Omega\,B_{\nu}(\nu,T_{\rm d})\,\left(1-e^{-\tau(\nu)}\right)$},
where $\Omega$\ is the solid angle of the emitting element,
$B_{\nu}$\ is the Planck function, $T_{\rm d}$\ is the
dust--temperature, and $\tau(\nu)$\ is the optical depth at frequency
$\nu$.  In a first iteration, we used a simple dust opacity model of
the form $\kappa_{\nu} \propto \nu^{\beta}$\ to fit for $\beta$, using
$\chi^2$\ minimization. We found that $\beta = 1.9$ -- $2$ fited the
data best over the entire cloud.  Note that compact structures were
smoothed out; therefore this cannot be taken as evidence against grain
growth in the centers of the two dense cores.  In a second step, we
then fixed the dust model to the \citet{ossen94} model for an MRN
grain size distribution, with thin ice mantles and no coagulation,
which has a $\kappa_{1.3\rm mm}=0.51$\,cm$^2$\,g$^{-1}$.  The
hydrogen--to--dust mass ratio was fixed to the canonical value of 110
\citep[e.g.,][]{sodroski97}.  We then searched for the optimum $T_{\rm
  d}$ and hydrogen column density $N_{\rm H}$ values (the two free
parameters) by calculating flux densities and $A_V$ values and then
comparing to the emission and extinction observations, using
$\chi^2$\ minimization.  The resulting line--of--sight--averaged
dust--temperature and column density maps are shown in
Fig.~\ref{fig:temp}, together with the SED fits at the central
positions of the two submillimeter peaks.  We found the best--fit
column--averaged dust--temperatures for the protostar and prestellar
core of $\sim\!17.7$~K and $\sim\!10.6$~K, respectively.  Despite the
fact that each individual image pixel was fitted independently and the
maps at different wavelengths have different outer boundaries, both
the temperature and the column density maps have very little noise and
are very smooth, demonstrating the robustness of our fitting approach.
Furthermore, we evaluated the effect of a possible filtered--out
extended emission component in the chopped 850~\micron\ and 1.3~mm
maps. Adding in offsets of 60\,mJy/15\arcsec\ beam at 850\,$\mu$m and
12\,mJy/11\arcsec\ beam at 1.3\,mm, corresponding to 20\% of the peak
surface brightness of the cold prestellar core, did not affect the
fitted temperature and column density (less than a 2\% change) for the
prestellar core or the protostar. Hence, we conclude that possible
missing extended emission in the (sub)mm maps does not affect our
results.

\section{Summary and conclusions}

As shown in Fig.~\ref{fig:temp}, the line--of--sight--averaged
dust--temperature decreases constantly without any significant jump
from $\sim\!17\pm1$~K at the outer boundary of the cloud, where
$N_{\rm H}$ drops below $10^{21}$\,cm$^{-2}$\ and $A_V \approx
0.3$\,mag, to $\sim\!13$\,K in the inner cloud at $N_{\rm H}\approx 1$
--- $2 \times 10^{22}$\,cm$^{-2}$\ and $A_V \approx 3$ -- $5$\,mag.
Note that due to the two major drawbacks of our simplified approach,
beam--smoothing and line--of--sight--averaging, the local
dust--temperatures at the source centers will be higher in the
protostar and lower in the prestellar core.  Nevertheless, the
boundaries of the two sources can be derived from the
dust--temperature map rather than from the smooth column density map.
We find that $T_{\rm d } = 13$\,K marks the mean inner
dust--temperature of the globule, from where the temperature towards
the protostar rises steeply and that of the prestellar core decreases.
We therefore adopt ellipses that follow the 13\,K contour in the
dust--temperature map as boundaries for the two sources (see Fig.~2).
The mean radii of these ellipses are 50\arcsec\ (10,000~AU) for the
protostellar core and 76\arcsec\ (15,000~AU) for the prestellar core.
Integration of the column density map within these boundaries yields
$M_{\rm H} = 1.6\pm0.3$\,M$_{\odot}$\ for the protostellar core and
$M_{\rm H} = 5\pm2$\,M$_{\odot}$\ for the prestellar core, where the
uncertainties are derived from an assumed $\pm 1$\,K uncertainty in
the dust--temperature.  The total mass of the globule (within the
$N_{\rm H} = 1\times 10^{21}$\,cm$^{-2}$\ contour) is $M_{\rm
  H}=15\pm5$\,M$_{\odot}$.  We note that $M_{\rm gas} = 1.36 M_{\rm
  H}$.  These masses imply that $\sim$45\% of the mass of the globule
is participating in the star--formation process.

Including the new \herschel\ data, we derive the following parameters
for the protostar: L$_{\rm bol}\!\sim\!1.5$~\lsun, L$_{\rm
  submm}$/L$_{\rm bol}\!\sim 4$\%, T$_{\rm bol}\!\sim\!62$~K, and
M$_{\rm env}$/L$_{\rm bol}\!\sim\!1.45$~\msun/\lsun,
\citep[cf.][]{laun10}, confirming the Class\,0 classification
according to \citet{chen95} and \citet{andre00}.  Furthermore, the
C$^{18}$O (2--1) FWHM line width for the prestellar core is
$\sim\!0.9$~km~s$^{-1}$ (Stutz et al., in prep.); using our fitted
prestellar core mass and temperature, we find that the ratio of
gravitational energy to thermal and turbulent energy is E$_{\rm grav}$
/ (E$_{\rm therm}$ $+$ E$_{\rm turb}) \simeq 0.9$, a marginally
sub--critical value which is expected for prestellar cores.  For
comparison, \citet{tobin10} derive a mass of $\sim$3 -- 4~\msun\ 
  for the prestellar core (integrated over a similar area and scaled
to a distance of 200~pc) using the 8~\micron\ shadow (see Fig.~1), a
good agreement given the uncertainties in both mass derivation
methods.  The prominent and extended 3.6~\micron\ coreshine (see
Fig.~1), originating from dust--grain scattering of the background
radiation field, is an indication of grain growth \citep{stein10} in
the \cb\ cloud.  These pieces of evidence together indicate that the
\cb\ globule, and other globules like it, are excellent sources in
which to study the earliest phases of low--mass star--formation.  As a
next step and in a follow--up paper, we will employ 3D--modeling to
overcome the effects of line--of--sight averaging and beam-smoothing,
and to reconstruct the full dust--temperature and density structure of
\cb\ and the other sources in our sample.  Measuring reliable
temperatures and column densities with the \herschel\ data in a sample
of prestellar and protostellar cores is a fundamental step towards
revealing the initial conditions of low--mass star--formation.

\acknowledgements The authors thank J.\ Tobin for helpful discussions
and D.\ Johnstone for a critical and helpful referee report.
PACS has been developed by a consortium of institutes led by MPE
(Germany) and including UVIE (Austria); KU Leuven, CSL, IMEC
(Belgium); CEA, LAM (France); MPIA (Germany); INAF- IFSI/OAA/OAP/OAT,
LENS, SISSA (Italy); IAC (Spain). This development has been supported
by the funding agencies BMVIT (Austria), ESA-PRODEX (Belgium),
CEA/CNES (France), DLR (Germany), ASI/INAF (Italy), and CICYT/MCYT
(Spain).  SPIRE has been developed by a consortium of institutes led
by Cardiff University (UK) and including Univ.  Lethbridge (Canada);
NAOC (China); CEA, LAM (France); IFSI, Univ. Padua (Italy); IAC
(Spain); Stockholm Observatory (Sweden); Imperial College London, RAL,
UCL-MSSL, UKATC, Univ. Sussex (UK); and Caltech, JPL, NHSC,
Univ. Colorado (USA). This development has been supported by national
funding agencies: CSA (Canada); NAOC (China); CEA, CNES, CNRS
(France); ASI (Italy); MCINN (Spain); Stockholm Observatory (Sweden);
STFC (UK); and NASA (USA).

\bibliographystyle{aa}
\bibliography{ms}

\end{document}